# Low-energy electronic interactions in ferrimagnetic $Sr_2CrReO_6$ thin films

Guillaume Marcaud[1*], Alex Taekyung Lee[1*], Adam J. Hauser[2], F. Y. Yang[3], Sangjae Lee[4], Diego Casa[5], Mary Upton[5], Thomas Gog[5], Kayahan Saritas[1], Yilin Wang[6], Mark P.M. Dean[6], Hua Zhou[5], Zhan Zhang[5], F. J. Walker[1], Ignace Jarrige[7], Sohrab Ismail-Beigi[1,4], Charles Ahn[1,4]

[1]Department of Applied Physics, Yale University, New Haven, Connecticut 06520, USA

[2]Department of Physics and Astronomy, University of Alabama, Tuscaloosa, Alabama 35487, USA

[3]Department of Physics, The Ohio State University, Columbus, Ohio 43210, USA

[4]Department of Physics, Yale University, New Haven, Connecticut 06520, USA

[5]Advanced Photon Source, Argonne National Laboratory, Argonne, Illinois 60439, USA

[6]Condensed Matter Physics and Materials Science Department, Brookhaven National Laboratory, Upton, New York 11973, USA

[7]National Synchrotron Light Source II, Brookhaven National Laboratory, Upton, New York, 11973, USA

**Abstract**

We reveal in this study the fundamental low energy landscape in the ferrimagnetic $Sr_2CrReO_6$ double perovskite and describe the underlying mechanisms responsible for the three low-energy excitations below 1.4 eV. Based on resonant inelastic x-ray scattering (RIXS) and magnetic dynamics calculations, and experiments collected from both $Sr_2CrReO_6$ powders and epitaxially strained thin films, we reveal a strong competition between spin-orbit coupling, Hund's coupling and the strain-induced tetragonal crystal field. We also demonstrate that a spin-flip process is at the origin of the lowest excitation at 200 meV, and we bring new insights into the predicted presence of orbital ordering in this material. We study the nature of the magnons through a combination of *ab initio* and spin-wave theory calculations, and show that two non-degenerate magnon bands exist and are dominated either by rhenium or chromium spins. The rhenium band is found to be flat at about 200meV (±25meV) through X-L-W-U high-symmetry points and dispersive toward Γ.

**Introduction**

Magnonics is an emerging alternative to modern electronics to carry and process information at high frequency and low-energy consumption using magnetic degrees of freedom. This field of research focuses on the use of spin waves or magnons in magnetic materials. This type of collective excitation is not limited by Joule heating as it does not involve the transport of electrons, and allows the superposition of signals through multiplexing. While ferromagnets have been first considered for this application, the community has recently turned to antiferromagnets for their higher frequencies of operation, the possibility of using left- and right-handed waves as an additional degree of freedom to carry information, and greater stability against parasitic magnetic fields. This stability results from the absence of net magnetization in antiferromagnets, which provides a relative immunity to external magnetic field [1] but creates challenges regarding excitation and control of spin waves.

The complex spin interaction in ferrimagnets leads to original spin dynamics that has been recently exploited to specifically overcome this obstacle [1–3]. While a promising ferrimagnet-based magnonic platform has already been demonstrated using $Y_3Fe_5O_{12}$ [4], other studies have revealed that unique spin characters can be realized in ferrimagnets due to the competition between different spins [1–3]. In rare earth–3d-transition metal ferrimagnetic compounds, fast field-driven antiferromagnetic spin dynamics are realized and field-driven domain wall mobility is remarkably enhanced at the angular momentum compensation temperature ($T_A$), where the net angular momentum vanishes [3]. At $T_A$, both the left- and right-handed spin waves intersect, and characters of antiferromagnetic and ferrimagnetic spin wave modes are observed [1].

Rhenium-based double perovskites (DP), $A_2B$ReO$_6$ ($A$=Sr, Ca, Ba, and $B$=Cr, Fe) are ferrimagnetic materials with a Curie temperature above room temperature [5]. They are strongly correlated materials, where the energy scales of spin-orbit coupling ($\lambda$), tetragonal crystal field ($\Delta_t$) and electronic correlations ($J_H$, U) compete, leading to a diverse manifold of ground states and phenomena, such as magnetic ordering above room temperature and an interplay between structural and electronic transitions [6]. Among the Re-based DPs, $Sr_2CrReO_6$ (SCRO) has the highest Curie temperature of $T_c$=620 K [7] (508 K from[8]) and shows a large strain-dependent magnetocrystalline anisotropy (MCA) [9].

Through a combination of state-of-the-art epitaxial growth of thin films, resonant x-ray scattering experiments and calculations, and spin-dynamics simulations, we explore in this work the mechanisms responsible for the fundamental low energy structure of SCRO and examine evidence for the presence of spin waves. We also bring additional insights into the presence of orbital ordering in this material [6]. The best agreement between theory and experiment reveals a strong competition between spin-orbit coupling, Hund's coupling and strain-induced tetragonal crystal field. We show that two non-degenerate magnon bands exist and are dominated by rhenium or chromium spins. The rhenium band is found to be flat at about 200meV (±25meV) through X-L-W-U high-symmetry points and dispersive toward Γ.

**Methods**

The three 90 nm (001) $Sr_2CrReO_6$ thin films characterized in this study were grown by off-axis magnetron sputtering on different substrates to achieve different strain states: -1% (compressive) on $(LaAlO_3)_{0.3}(Sr_2AlTaO_6)_{0.7}$ (LSAT), +1% (tensile) on a relaxed $SrCr_{0.5}Nb_{0.5}O_3$ (SCNO) and 0% (unstrained) on $SrTiO_3$ (STO). No additional phases are detected through x-ray diffraction experiments, and the quality of the epitaxy is confirmed with the (103) reciprocal space map presented Figure S1 of the supplementary material [10].

The RIXS characterization was performed at the Advanced Photon Source at the 27-ID beamline. The incident photon energy was selected by a high-resolution mmE Si(440) monochromator near the Re-$L_3$ edge (10.539 keV) and the final photon energy was analyzed with a 2 meter Si(119), leading to an overall resolution (Full Width at Half Maximum - FWHM) of 70 meV in this configuration. With an analyzer of 10 cm diameter 200 cm away from the sample, the momentum resolution in h, k and l is estimated to be lower than 0.07 r.l.u. in the pseudocubic system of coordinates. To enhance the signal coming from the films and to reduce the elastic background intensity, the incident beam grazes the sample surface (0.5 < θ < 9 degrees for thin films, depending on the vector q, and 1.2 degree for the powder) while the scattering angle 2θ was kept close to 90 degrees in a horizontal scattering geometry.

Density functional theory (DFT) calculations are performed using projector augmented wave (PAW) method [11] implemented in the VASP code [12]. We use revised version of the generalized gradient approximation (GGA) proposed by Perdew et al. (PBEsol) [13]. Spin-orbit coupling (SOC) is not included, since the effect of SOC is weak for SCRO from our previous study [6]. We use 500 eV of kinetic energy cutoff and 9×9×7 of k-point mesh.

The RIXS calculations were performed using EDRIXS, an open-source toolkit for simulating XAS and RIXS spectra based on exact diagonalization of model Hamiltonians. It is developed as part of the COMSCOPE project in the Center for Computational Material Spectroscopy and Design, Brookhaven National Laboratory [14]. We use a single atomic multiplet model, considering a single $Re^{5+}$ ($d^2$) ion. We obtain the crystal field splitting parameters from the Wannier function projections on Re d states [15], and use these parameters as an initial guess. Then we explored a large phase space of parameters near the set of initial values. We fix the $e_g$-$t_{2g}$ splitting (10Dq) of 3.23 eV from Wannier projection, since 10Dq is large and thus only $t_{2g}$ orbitals are important on the RIXS spectra below 1.5 eV.

The magnon spectrum are computed via DFT and spin-wave theory following Toth et. al [16]. The exchange coupling parameters $J$ of a Heisenberg model are obtained from DFT, and we assume an isotropic magnetic interaction $J_x=J_y=J_z$. We compute the energy difference between the ferromagnetic (FM) and antiferromagnetic (AFM) configurations $\Delta E = E[\text{AFM}] - E[\text{FM}] = \sum_{i,j} J \boldsymbol{S}_i \cdot \boldsymbol{S}_j$, where $\boldsymbol{S}_i$ and $\boldsymbol{S}_j$ are spins of Cr ($S$= 3/2) and Re ($S$= 1/2) and obtain $J$ = 33.58 meV. We then use this Heisenberg model to compute the magnon spectrum via spin-wave theory. The main equations are given in the supplementary material [10].

The REXS experiments were carried out at the Advanced Photon Source at the 33-ID beamline with a '4S+2D' six-circle diffractometer. The data are collected with a Pilatus solid state detector located about 1 meter from the SCRO samples, which are cooled to a temperature of 15 K under a beryllium hemispheric dome. The expected reflection induced by the orbital-ordering (OO) at (0 0 ½) is reached with an incident angle $θ_i$=4 degrees and measured in a fixed q mode as a function of energy in the 10.5 to 10.6 keV range, with azimuthal angle ɸ defined as the angle between the <100> crystal axis of SCRO and the in-plane component of the incident beam. The polarization sigma of the incident beam is fully in-plane. The fluorescence spectra are extracted from the background signal of the detector during the energy scans and the (0 0 ½) reflections are captured by a smaller region of interest and after background subtraction. The energy-dependent REXS intensity of the OO-induced reflection is calculated from the DFT-derived atomic positions of Sr, Cr and O in a doubled SCRO unit cell (under tensile, compressive, or relaxed strain), the complex scattering factor extracted from the absorption (XAS) spectrum that result from the RIXS calculations in the presence of orbital-ordering, and the Cromer-Mann coefficients. The intensity is then finally corrected for self-absorption. See the supplementary material for more details[10].

**Results**

An overview of the energy scales in the RIXS map of SCRO powder near the Re-$L_3$ edge is presented in Figure 1. In addition to the elastic line at 0 eV, sharp excitations near 1 eV are observed (zone I), as well as broader features at 5, 8 and 9-10 eV (zone II, III, and IV, respectively). The resonance of some of these excitations at specific energies of the incident light provides a signature of the electronic states involved in the intermediate $|n\rangle$ and final $|f\rangle$ steps of the RIXS process [17], as described schematically in Figure 1 a). The excitations in zone I are enhanced when the incident photon energy is set to $E_i$=10.534 keV, while zone II resonates at $E_i$=10.538 keV. We attribute this difference of 4 eV to the splitting of the $t_{2g}$ and $e_g$ levels of rhenium by an octahedral crystal field, similar to what is observed in another rhenium-based double perovskite $Ba_2YReO_6$ [18]. The broad zone III dominates the spectrum in the incident energy range between the resonances of zone I and II, and is identified as a charge transfer from the rhenium to the oxygen ligand, whereas zone IV features a linear dependence with incident photon energy and is attributed to fluorescence. While the excitations in zone I only involve electronic transitions from and to the $t_{2g}$ levels of rhenium, namely *dd intra-$t_{2g}$*, we identified the excitations in zone II as the result of transitions from $t_{2g}$ and to $e_g$, namely *dd $t_{2g}$-$e_g$*, although other excitations yielding broad spectra, like inter-band or charge-transfer excitations, cannot be excluded. A closer look at the spectral weight distribution in zone I reveals three components. The two higher energy peaks at 0.55 and 0.9 eV are likely *dd intra-$t_{2g}$*, further split by strain, spin-orbit coupling, or Hund's coupling, but a feature as low as 0.2 eV suggests the presence of an additional relaxation channel such as bosonic excitations or the suspected orbital ordering (OO) in SCRO, as discussed in a recent review on rhenium-based double perovskites [6].

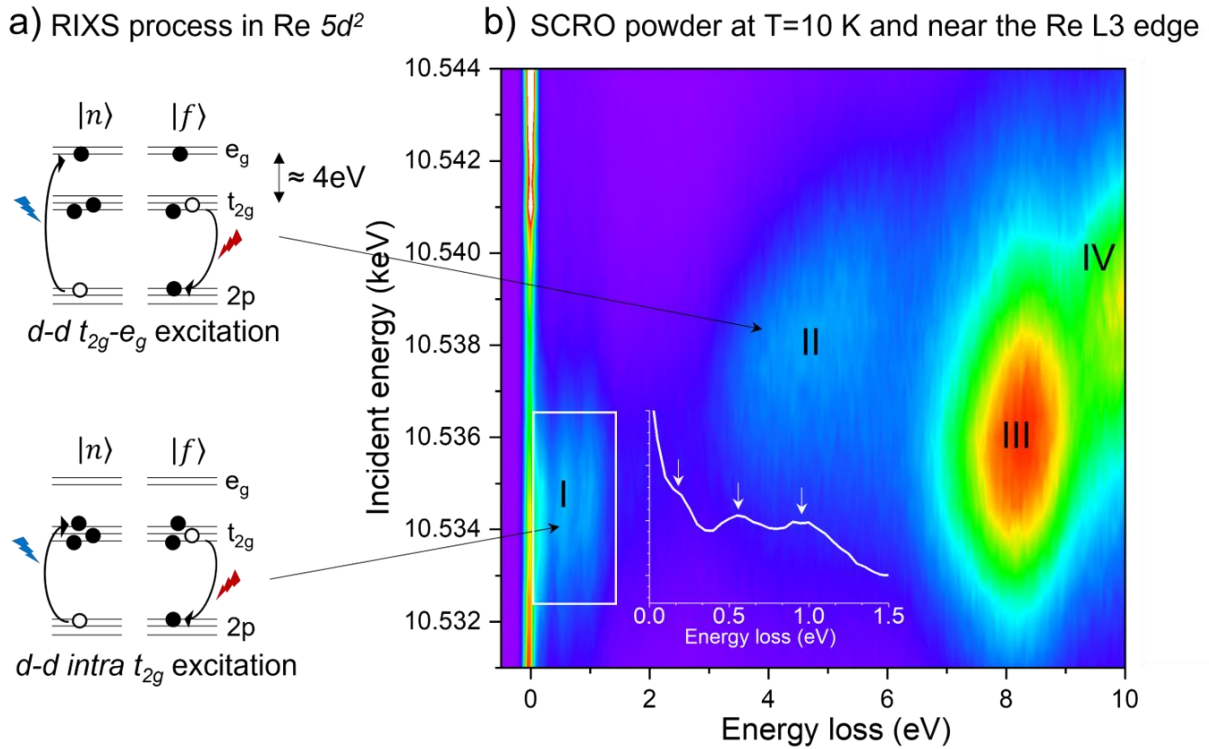

*Figure 1: **Incident photon energy versus energy loss RIXS mapping**, near the $L_3$-edge of rhenium for a powder of SCRO at T=10 K. In addition to the elastic line at 0 eV, four zones (I-IV) are identified. I) The lowest energy part of the spectrum (energy loss<1.4 eV) includes d-d transitions between the $t_{2g}$ orbitals of rhenium, namely intra-$t_{2g}$, and potential quasiparticle excitations. II) The*

*broader feature between 4 and 6 eV resonates 4 eV higher than the ones in zone I. Its spectral weight is identified as d-d transitions between the $e_g$ and $t_{2g}$ orbitals. The strong signal located at 8 eV results from transitions between oxygen 2p and rhenium 5d (charge transfer) and the linearly incident-energy-dependent intensity in zone IV is induced by fluorescence.*

Recent progress in the epitaxial thin film growth of SCRO on different substrates [9, 19, 20] allows one to investigate the momentum dependence of these excitations and the role of strain on the low-energy excitations below 1.4 eV (zone I). The data in Figures 2, 3 and 4 have been collected from 90 nm thick SCRO thin films under tensile (+1%) and compressive strain (-1%).

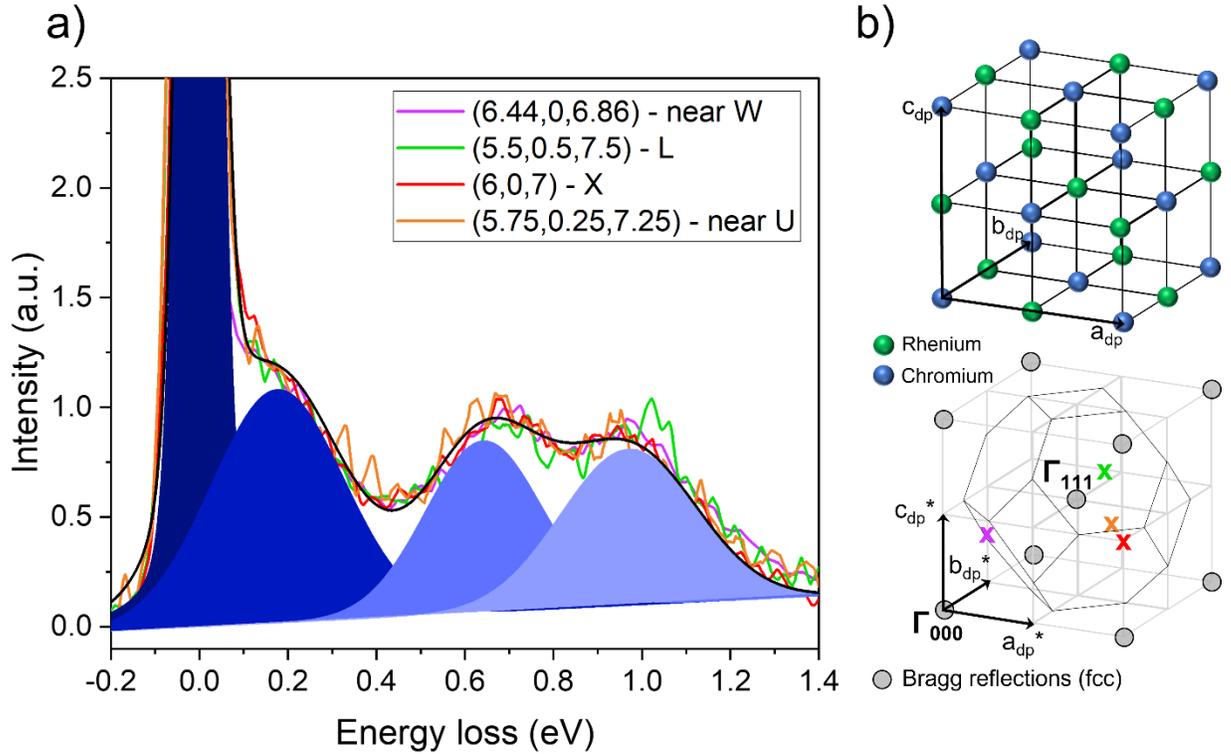

*Figure 2: **RIXS spectra for different momentum transfer**. a) Low-energy RIXS spectra (zone I in the map of Figure 1) collected from the compressive SCRO thin film for different momentum transfer q expressed in the double perovskite cubic system (dp), at a temperature of 17 K. The spectra are normalized by the integrated intensity between 0.05 and 1.4 eV. Three excitations are resolved at 0.2, 0.65 and 1 eV, in addition to the elastic line at 0 eV. The black curve represents the sum of four Voigt functions that are used to fit the experimental data collected at the X point. b) Schematic unit cell of the fcc- SCRO in real space, and the corresponding reciprocal space and Brillouin zone. The X marks represent the momenta that have been probed, with the same color code used for the spectrum.*

Three non-elastic features are clearly resolved in Figure 2a) at 0.2, 0.65 and 1 eV for the SCRO film under compressive strain and at a temperature of T=17 K. We do not observe significant dispersion of these excitations for all measured momentum transfers, at L, X, and near U and W high-symmetry points. The Γ point could not be measured because of the Bragg reflection that occurred at this coordinate (see figure 2b), which would dominate the spectrum with a strong elastic peak centered at 0 eV and prevent the observation of non-elastic features below 1.4 eV. We also note that our RIXS measurements required a grazing incidence condition of the incident photons to maximize the interaction of the light with the material and improve the signal to noise ratio, which limits our capability to navigate through reciprocal space, specifically W and U.

The absence of dispersion in Figure 2, also observed in an unstrained (0% strain) SCRO film as shown in Figure S2 of the supplementary material[10], is consistent with a local *d-d intra-t$_{2g}$* mechanism but remains surprising in this energy window below 0.5 eV. Similar low-energy excitations have been reported in powder samples of $Ca_2FeReO_6$, $Ba_2YReO_6$, $Ba_2FeReO_6$ and in the iridate $Sr_3CuIrO_6$, but their origin is unclear. They have been ascribed to multi-phonon excitations in the rhenates [18] or impurities in iridates [21]. The former is unlikely at the high energy of the Re L$_3$-edge because of the very short core-hole lifetime that reduces the RIXS cross-section for phonons [22–24].

Next, we delve deeper into the role of strain in SCRO. The direct comparison of the RIXS spectra of two films under opposite and equal strain, Figure 3 a) and b), directly reflects the effect of a crystal-field sign change on the electronic structure. The elastic line has been removed to better resolve the peak at 0.2 eV. The three excitations discussed previously are observed for all temperatures in the 17-300 K range and for both strain states. From compressive to tensile strain, the positions of the excitations shift slightly toward lower energies and the spectral weight is transferred from the feature near 1 eV to the one at 0.2 eV. The three excitations are independent of temperature for the compressive SCRO film, but a temperature dependence arises for T>100K when the strain is reversed to tensile, where the excitation at 0.2 eV weakens in favor of the other two excitations.

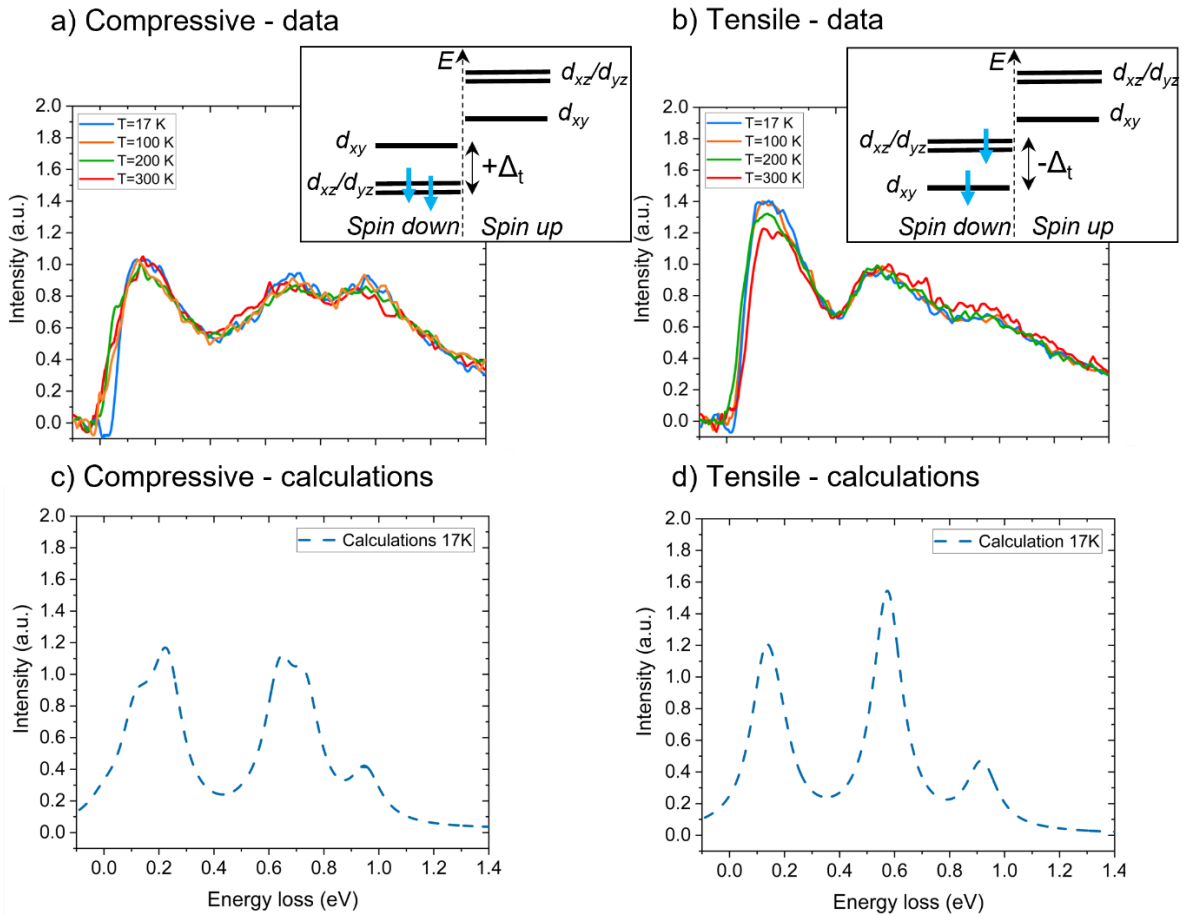

*Figure 3: **RIXS spectra collected at different temperature** for a) compressive SCRO film at q=(6.4, 0.34, 6.8) and b) tensile SCRO film at q=(6, 0.34, 7.2) in the 17-300 K range. The elastic line at 0 eV has been removed for a better reading of the low energy excitations in the experimental data, and the spectra are normalized by the integrated intensity between 0.05 and 1.4 eV. c) and*

*d) Best-matched EDRIX calculations at 17 K in presence of an exchange magnetic field. The calculated intermediate and final state allowed us to identify the peak near 0.2 eV as the result of a spin-flip process. The insets show the electronic level splitting due to the Hund's coupling and strain-induced tetragonal crystal field ($\Delta_t$).*

A few scenarios can be considered to explain the emergence of these low-energy excitations. The first one is the effect of spin-orbit coupling, Hund's coupling and epitaxial strain. While spin-orbit coupling can further split the $t_{2g}$ levels into a $J_{eff}$=3/2 and 1/2 subset, this by itself is not sufficient to explain the three excitations in the experimental data, but spin-orbit combined with a crystal field induced by the epitaxial strain ($\Delta_t$) may result in three excitations. We investigated this idea using an atomic multiplet model solved by exact-diagonalization, as implemented in the EDRIXS software [14]. We systematically explored a range of values of the on-site Coulomb interaction (U), spin-orbit coupling ($\lambda$), Hund's coupling ($J_H$) and tetragonal crystal field ($\Delta_t$) induced by the strain, in the range $0 \leq \lambda \leq 0.4$, $|\Delta_t| \leq 0.5$, $0.1 \leq J_H \leq 0.6$ and U=0, 2 (unit of eV), and typical results are presented Figure 4 a) and b). No good match with the experimental data was obtained for the two states of strain, especially for the lowest energy excitation, ruling out this first mechanism.

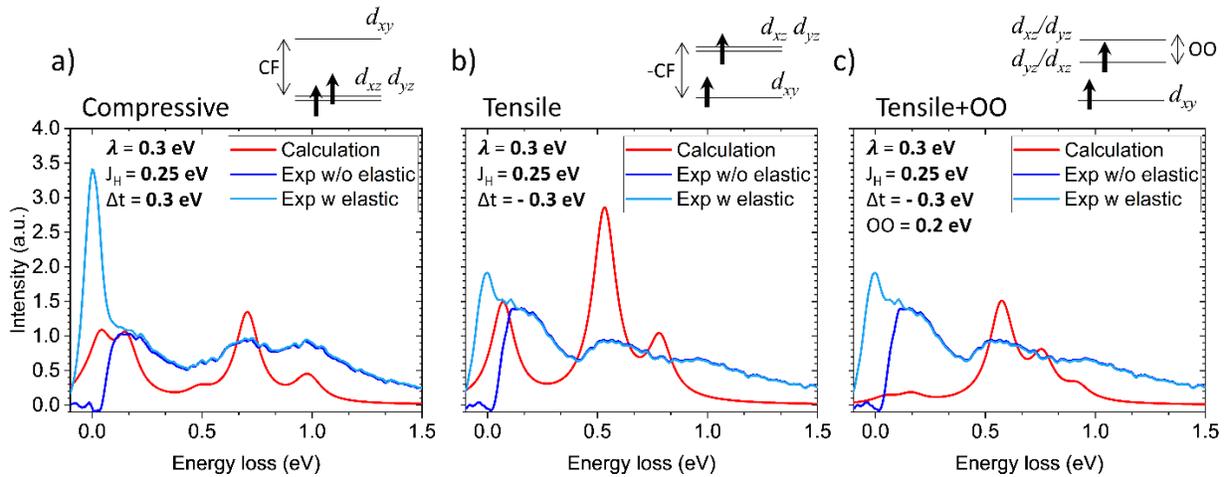

*Figure 4: RIXS calculations and experimental data for (a) compressive and (b)-(c) tensile strained SCRO at T=17 K. A large set of input parameters, $0.2 \leq \lambda \leq 0.4$ eV, $-0.4 \leq \Delta_t \leq 0.4$ eV, $0 \leq OO \leq 0.2$ eV, and U=0, 2 eV, are used to match the experimental data. The best set is found to be U=2, $\lambda = \Delta_t$ =0.3 eV and $J_h$=0.25 eV, although the quality of the fit is poor, especially for the excitation at 0.2 eV for tensile SCRO. The agreement between experiment and calculation is not improved with OO as shown in (c).*

The second scenario that we considered is based on the presence of orbital ordering. A recent theoretical study showed using DFT+U calculations that SCRO may stabilize at low temperatures in a new phase where the Re $d_{yz}$ and $d_{xz}$ orbitals spontaneously order in a checkerboard fashion, which results in an energy gap formation [6]. The presence of a 0.2 eV gap has also been claimed in another experimental study based on IR-absorption spectroscopy data [8] and might share a common origin with the 0.2 eV excitation observed here in RIXS. Using the same atomic model, we mimicked the effect of OO on the energy levels of rhenium with a site-dependent splitting of the $d_{yz}$ and $d_{xz}$ orbitals via an additional OO energy parameter in the calculation, averaging the two Re sites involved in the checkerboard-like orbital ordering described in reference [6]. We explored the range $0 \leq OO \leq 0.2$ eV and obtain the lowest energy peak near 0.2 eV with OO=0.2 eV, as shown in Figure 4 c). However, the relative magnitude of the peak is too small, indicating that OO may not be the origin of the lowest energy excitation near 0.2eV.

Based on this result, we performed an additional experimental characterization on these SCRO thin films, which aimed at revealing the OO by Resonant Elastic X-ray Spectroscopy (REXS). Since OO slightly modifies the absorption spectrum (XAS) of the two types of rhenium involved in the orbital ordering, new Bragg reflections at specific reciprocal space coordinates should be observed. We note that due to the nature of the orbitals involved in the predicted OO ($d_{yz}$ and $d_{xz}$), the intensities of the emergent Bragg reflections not only depend on the incident photon energy near the Re $L_3$ edge, but also on the photon polarization direction relative to the orbitals as shown from the calculation Figure 5 a). We experimentally probed these new Bragg positions as a function of the photon energy near the Re $L_3$ edge and for different polarization directions, and compared the results to REXS calculations for the tensile SCRO in Figures 5 b). Same results are obtained for the compressive SCRO in the supplementary material Figure S4 [10]. No additional Bragg reflections are observed in this REXS study, from which we conclude that there is no long-range OO in our SCRO. Impurities, defects, or strain gradients in SCRO (see Figure S1c) may be responsible for the absence of long-range order.

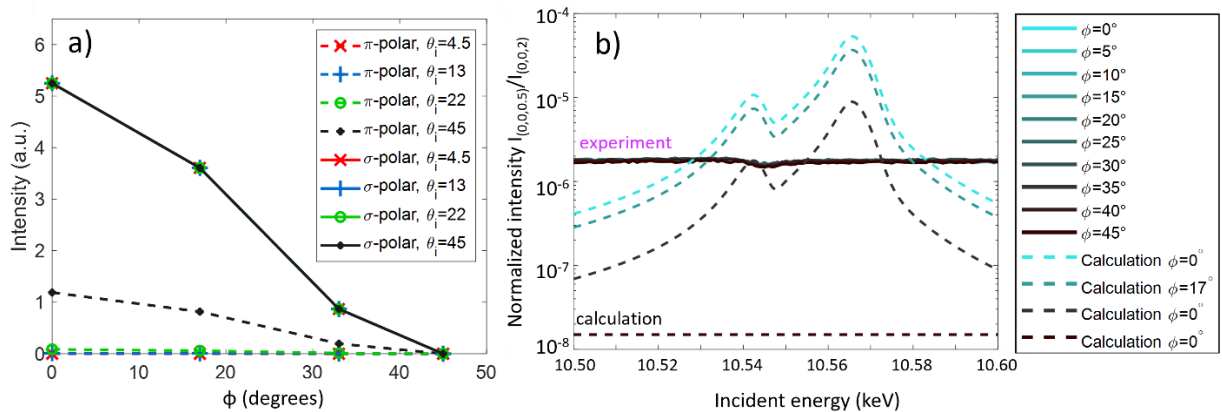

*Figure 5: Experimental and calculated energy scans at q=(0 0 ½) for the OO-induced reflections in tensile SCRO. a) Calculated intensity of the OO-induced reflection at (0 0 ½) and an incident energy $E_i$=10.565 keV, as a function of the azimuthal angle φ, for different incident angle $θ_i$ and for π, σ polarization. b) Experimental (solid lines) and calculated (dashed lines) energy-dependent intensity of the OO-induced reflection for σ polarization, $ϑ_i$=8.5 degrees and for different φ values. The simulation is calculated from the absorption spectra resulting from the EDRIXS calculation with OO for the tensile SCRO. The data are corrected for self-absorption and normalized by the Bragg reflection intensity at q=(0 0 2).*

A third scenario is the presence of magnons. While their dispersion has not been observed in Figure 2, they remain a promising candidate in this low-energy part of the spectrum, especially in a ferrimagnetic material such as SCRO. One way to distinguish magnons from charge excitations experimentally is to use a more elaborate spectrometer to characterize the changes in the polarization between the incident and outgoing x-rays in RIXS [25]. Here, we investigated theoretically the presence of an exchange magnetic field in the range 0 ≤ M ≤ 0.1 eV in our atomic model. We also took into account the direction of this field by considering the strain-dependent magnetic anisotropy in SCRO thin films [9]. The magnetic easy axis is perpendicular (parallel) to the film surface when the strain is tensile (compressive), which translates into a field along the z-direction [0, 0, M] (in the xy-plane $[M/\sqrt{2}, M/\sqrt{2}, 0]$) in the model. We obtained good agreement between theory and experiment in this framework, as shown in Figure 3 c) and d). We use $λ$=0.3, $J_h$=0.25, U=2, $Δ_t$ = −0.2, and M=[0, 0, 0.05] for tensile strain, and $λ$=0.3, $J_h$=0.15, U=2, $Δ_t$ = 0.3, and M=[0.035, 0.035, 0] (unit of eV) for compressive strain. The discrepancy in the peak intensities is likely due to the limitation of the single atomic model, which excludes any form of interaction between the Re and the surrounding ions. Each peak originates from many excitations that are close in energy and that are broadened by a core-hole lifetime of 0.065 eV [26]. Using the calculated intermediate and final states, we can identify the 0.2 eV excitation as a spin-flip process for both compressive and tensile strain, which is allowed by symmetry [27]. For example, spin flip from $d_{xz}^\downarrow/d_{yz}^\downarrow$ to $d_{xz}^\uparrow/d_{yz}^\uparrow$ is allowed for compressive strain.

The decrease in intensity observed for the low-energy peaks in tensile SCRO at high temperatures is still an open question. While it is known that bulk SCRO undergoes a transition from a tetragonal to a cubic phase at 260 K [7], this transition will be altered in the thin film due to the clamping effect of the substrate. To estimate the effect of crystal field, we provide a summary of the effect of the tetragonal crystal field parameter |Δt| in Figure S3, and demonstrate that the lowest energy peak is shifted towards lower energy levels, resulting in a reduction in magnitude near 0.2eV after eliminating the elastic peak. However, the peak near 0.6eV is also shifted to ~0.4 eV, which is inconsistent with the experiment. Future studies are needed to understand the temperature dependence of the low-energy peaks depending on the strain.

While the physics of a spin-flip process can be understood by a single Re-atom model, it cannot describe the dispersion of the spin excitation (magnon) in reciprocal space, which is typically large in antiferromagnetic cuprates or iridates [28–31], nor can it describe the spin-spin interaction between Cr and Re. We address this challenge by combining DFT calculations and spin-wave theory [16]. The exchange coupling parameters J of a nearest-neighbor Heisenberg model are obtained from DFT, and we assume an isotropic magnetic interaction $J_x=J_y=J_z$. We did not include SOC for DFT calculations because the effect of SOC is almost negligible [6]; due to the large magnetic moments and strong magnetic interaction in SCRO, the effect of SOC is quenched within DFT. We extract the nearest neighbor Heisenberg coupling $J$ = 33.58 meV (see method section), and use the Heisenberg model to compute the magnon spectra. We obtain $T_C$ = 415 K, which is comparable to the experimental value of $T_C$ = 508 K from our fully ordered film [8]. The discrepancy between the spin-wave model and the experiment may be due to the next nearest neighbor interaction: previous theoretical work on $Sr_2CrOsO_6$ predicts nearest ($J_1$) and next nearest neighbor ($J_2$) interaction parameters of $J_1$=35meV and $J_2/J_1$=0.4, and the predicted $T_C$ is 725 K that includes spin canting [32].

As presented in Figure 6, there are two magnon bands whose energies are 0 and 110 meV at Γ, and 201 and 302 meV at X. Differing from typical antiferromagnetic materials like cuprates and iridates [30], which have doubly degenerate magnon bands because of equal up- and down-spins magnitudes ($|S_A|$ = $|S_B|$, where A and B are the two opposite-spin TM ions) [33], SCRO is ferrimagnetic and the two lowest magnons bands are not degenerate ($|S_{Cr}|$ > $|S_{Re}|$). We also note that the dispersion is only about 25 meV for both bands along the X-L-W-U path but reaches 200 meV between Γ and any of these high-symmetry points.

Unlike antiferromagnetic materials, the different TM ions with majority spin and minority spin can be disentangled in the magnon band in SCRO. In Figure 6, we also present the relative weight of the Re spins in magnon bands [16, 34, 35] (see supplementary material [10] for details). We find that the lower magnon band is Re-dominated and the upper one is Cr-dominated, and that this differentiation is largest at the k-points far from Γ. Specifically, along the X-L-W-U path, the lowest magnon band at 200 meV is almost exclusively from rhenium. For this reason, and because of the large energy difference between the absorption spectrum (XAS) of Cr and Re, only Re-dominated bands are observed in a RIXS experiment at the Re $L_3$ edge at the wave vectors considered in our experiments. The energy of this experimental Re-band is about 200 meV along the X-L-W-U path, in agreement with our theory. This finding shows that RIXS can be used to disentangle the majority and minority spins from different TM ions such as 3d-4d or 3d-5d ferrimagnetic double perovskites $A_2BB'O_6$ (A=Sr, Ca, B=Cr, Fe, B'=Mo, W, Re).

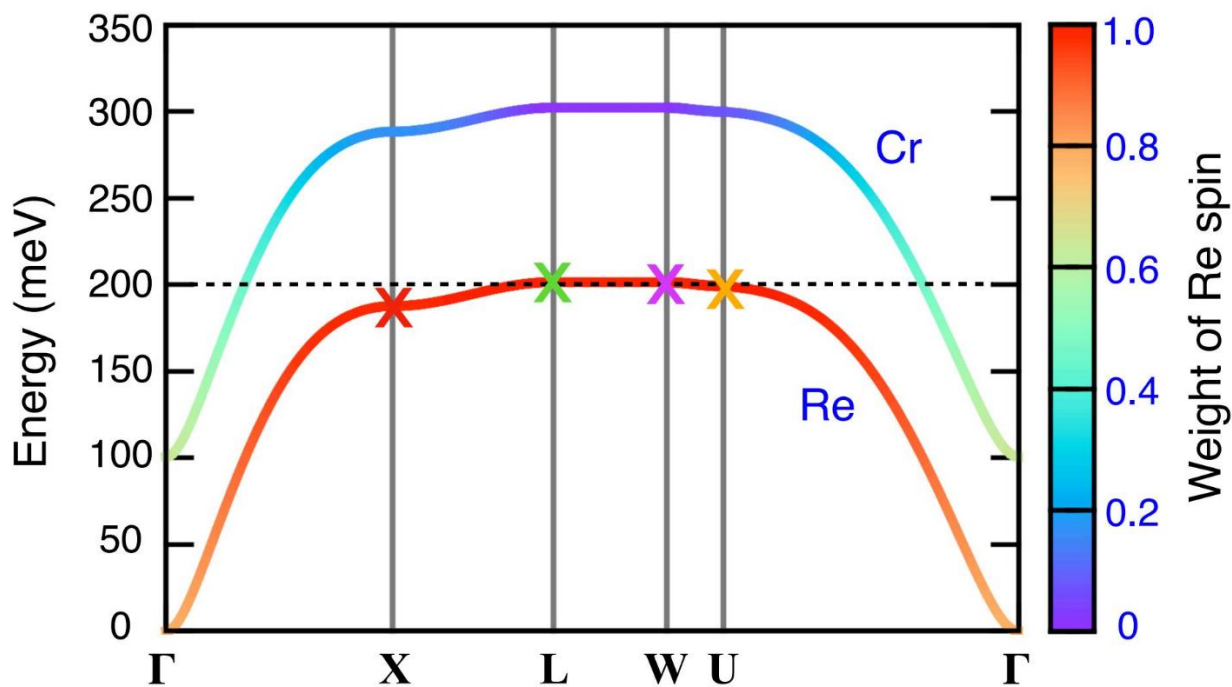

*Figure 6: Magnon band structure. Two lowest, non-degenerate magnon bands. Color scale indicates the weight of Re spin involved in the magnon mode.  Exchange parameters are obtained from the DFT and spin-wave calculation. The X markers represent the experimentally measured k-points and their energies.*

**Conclusion**

We identified in this study the origin of the three excitations observed in the RIXS spectra below 1.4 eV for a $Sr_2CrReO_6$ powder and thin films under different states of strain. The strain mainly leads to a spectral weight redistribution below 1.4 eV and a temperature dependence of the lowest energy peak at 0.2 eV for tensile strain. Three scenarios were explored by means of RIXS calculations to understand the experimental data, based on (i) a combination of spin-orbit coupling, Hund's coupling and strain-induced tetragonal crystal field, (ii) an orbital ordering between the $d_{yz}$ and $d_{xz}$ orbitals of rhenium, as predicted in a recent theoretical study, and (iii) the presence of magnons in $Sr_2CrReO_6$. A good match between theory and experiment was obtained in the second (ii) framework but subsequent REXS experiment experiments showed no trace of long-range OO-induced Bragg reflections, which may be due to strain gradients or the presence of defects in the strained $Sr_2CrReO_6$ films.  This leads us to consider scenario (iii).

We showed that a spin-flip process, mainly on the Re site, could also be responsible for the lowest 0.2 eV excitations, and that the limited dispersive behavior observed in experiment is fully compatible with the magnon calculations. The best fit of the RIXS spectra for the different strain states within the magnon theory framework is obtained with similar values of spin-orbit coupling, Hund's coupling and strain-induced tetragonal crystal field, which also emphasizes the competitive nature of the energy scale in $Sr_2CrReO_6$, which is likely responsible for the rich physics observed in this compound.

**ACKNOWLEDGMENTS**

Work at Yale was supported by the Air Force Office of Scientific Research (AFOSR) under Grant No. FA9550-21-1-0173.

Work at Brookhaven National Laboratory was supported by the U.S. Department of Energy, Office of Science, Office of Basic Energy Sciences under Contract No. DE-SC0012704.

This research used resources of the Advanced Photon Source, a U.S. Department of Energy (DOE) Office of Science user facility operated for the DOE Office of Science by Argonne National Laboratory under Contract No. DE-AC02-06CH11357.

F.Y.Y. Acknowledges the support by the Center for Emergent Materials, an NSF MRSEC, under Grant No. DMR-2011876.

*Alex Taekyung Lee and Guillaume Marcaud contributed equally to this work.